\documentclass[
    ,final            
   ,numberedheadings 
	,letterpaper
  ]
  {aipproc}

\layoutstyle{8x11double}

\usepackage{amssymb}

\begin{document}

\title{Spectral trends in the Very High Energy blazar sample \newline due to EBL absorption}


\author{Bagmeet Behera}{
  address={Landessternwarte, Universit\"{a}t Heidelberg, K\"{o}nigstuhl, D69117 Heidelberg, Germany}
}

\author{Stefan J. Wagner}{
  address={Landessternwarte, Universit\"{a}t Heidelberg, K\"{o}nigstuhl, D69117 Heidelberg, Germany}
}


\begin{abstract}
The absorption of $\gamma$-rays in the intergalactic medium due to the EBL (Extragalactic Background Light) causes the observed blazar spectrum to be fainter and softer than their intrinsic state. It could thus be expected to see an effective spectral-softening trend with redshift. No such trend is evident in the sample of VHE blazars currently observed.

To check which distributions of the properties of the parent blazar population could reproduce the observations, various simulations are done. The resulting subsamples that satisfy a generic detection criterion for the current generation of ACTs (Atmospheric Cherenkov Telescope) are checked to identify whether any inherent correlations (of spectral properties with redshift) are required to explain the current observations.
		\end{abstract}
\maketitle
\section{Introduction}
Advances in ground based ACT (Atmospheric Cherenkov Telescope) techniques in the last two decades has resulted in a dramatic increase in the number of extragalactic objects detected in VHE (Very High Energy, defined as E\,$> 100$\,GeV) $\gamma$\,-rays. The VHE blazar sample has increased from just a few, to over 20 at present. Most of the detected AGN are classified as HBLs (High frequency peaked Bl-Lac object); however, two LBLs (Low frequency peaked Bl-Lac object), one nearby FR-I (Fanaroff Riley type-I) radio galaxy, and one FSRQ (Flat Spectrum Radio Quasar) have now been detected. The maximum redshift (z) of a source detected with high significance is 0.536 for 3C279. However, this source has been detected only within a one day period, from all observations made public. The redshift (confirmed) of the farthest, relatively steady source is 0.212 for 1ES\,1011+496. Thus the AGN zoo is now open for study in the highest energy regime.

VHE emission from all extragalactic sources suffer extinction in the intergalactic medium due to the EBL (Extragalactic Background Light). This extinction is the result of pair-production from photon-photon interactions ($\gamma_{VHE} + \gamma_{EBL} \rightarrow e^+ + e^-$), having the most effective cross-section for the product of photon energies \mbox{$\approx1$\,TeV\,$\times1$\,eV} (i.e. the VHE photon energy taken in TeV and the EBL photon energy taken in eV). The EBL in the UV-optical and the IR band provide enough photon density to get noticeable absorption of source photons in energies $\gtrapprox 100$\,GeV, and this absorption increases sharply above a few TeV. The net optical depth ($\tau$) in this process is naturally dependent on z, giving $\tau = \tau(z,E_{VHE})$. The absorption of $\gamma$-rays, hence causes the observed blazar spectra to be softer than their intrinsic state, with a more pronounced effect for large-z sources. 

In this work, the VHE blazar sample is studied for any discernable trends in the intrinsic and observed spectral features with z. It has been attempted to explain the spectral trends or the lack thereof, by simulating parent samples with various intrinsic properties, that are either derived from the observed sample, or assumed from certain blazar-unification models.

\section{VHE blazar sample}
The VHE blazar sample as known at the time of this symposium is described in Table \ref{rankingtable}. Columns $3$ and $4$ are the observed photon index and the normalisation at $1$\,TeV to a simple power law fit ($\frac{dN}{dE} = N_{0,fit} E^{-\Gamma_{obs}}$), respectively. The fit to the observed data was converted into an intrinsic spectra by correcting for the attenuation due to the EBL; the upper limits given in \cite{Aharonian2006a} was the template EBL used for this work.  The slope of the corresponding deabsorbed spectra (calculated by taking a straight line fit between $0.2$\,TeV and $1$\,TeV) is given in column $6$, is the intrinsic photon index. The table is sorted by redshift. It is clear from this table as well as Figure \ref{Simulation1n2} and Figure \ref{Simulation3} (squares) that there is no evidence of a spectral trend with z, for both the intrinsic and the observed photon indices.
\begin{table}[!!h!!t]
\centering         
\caption{The VHE blazar sample, sorted by redshift. The values for the observed photon indices and the normalization (to a power law fit) at $1$\,TeV are from the references in the last column. The Integral flux in column 5 is calculated from the power law fit with the given $\Gamma_{obs}$ and normalization, in columns 3 and 4 respectively. The 6th and 7th column give the approximate calculated values (corrected for EBL absorption) for the intrinsic $\Gamma$ and the deabsorbed integral flux between $0.2$\,TeV to $10$\,TeV.}
\label{rankingtable}                           
\begin{tabular}{llccccclll}\hline\hline
 (1) & (2) & (3)         & (4)                      &  (5)        &  (6)           & (7)         &  (8)       &   (9)  \\
Name & z & $\Gamma_{obs}$&  N$_{0,fit}$(at$1$\,TeV) & F$_{obs}$   & $\Gamma_{int}$ & F$_{int}$\,=\,L$_{TeV}$   & Instrument & Ref.\\
     &   &               &  $\times 10^{-13}$       & $\times 10^{-11}$  &                & $\times 10^{-11}$  & (for spectra) &   \\
     &   &               &  [/cm$^2$/s/TeV]         & [/cm$^2$/s] &                & [/cm$^2$/s] &            & \\\hline  
Mrk 421 & 0.031 & 3.19 & 380.1 & 58.9 & 3.0 & 65.2 & HEGRA & \cite{Aharonian2002} \\ 
Mrk 501 & 0.034 & 2.76 & 84.0 & 8.10 & 2.5 & 9.30 & HEGRA & \cite{Aharonian2001} \\
1ES 2344+514 & 0.044 & 2.95 & 15.5 & 1.83 & 2.7 & 2.17 & MAGIC & \cite{Albert2007a} \\
Mrk 180 & 0.045 & 3.25 & 8.99 & 1.49 & 3.0 & 1.74 & MAGIC & \cite{Albert2007b}  \\
1ES 1959+650 & 0.048 & 2.72 & 43.0 & 3.98 & 2.4 & 4.89 & MAGIC & \cite{Albert2006a}\\
Bl Lacertae & 0.069 & 3.64 & 2.37 & 0.63 & 3.2 & 0.784 & MAGIC & \cite{Albert2007c}\\
PKS 2005-489 & 0.071 & 4.0 & 1.66 & 0.69 & 3.5 & 0.851 & HESS & \cite{Aharonian2005a}\\
RGB J0152+017 &  0.08 & 3.53 & 4.40 & 1.02 & 3.0 & 1.34 & HESS & \cite{Aharonian2003}\\
PKS 2155-304 & 0.116 & 3.32 & 20.0 & 3.61 & 2.6 & 5.851 & HESS & \cite{Aharonian2005b}\\
H1426+428 & 0.129 & 3.55 & 185.0 & 43.9 & 2.7 & 73.7 & Whipple & \cite{Horan2002}\\
1ES 0229+20 & 0.14 & 2.5 & 6.23 & 0.463 & 1.6 & 1.20 & HESS & \cite{Aharonian2007d} \\
H 2356-309 & 0.165 & 3.06 & 3.08 & 0.412 & 2.0 & 1.03 & HESS & \cite{Aharonian2006a}\\
1ES 1218+304 & 0.182 & 3.0 & 101.0 & 12.7 & 1.7 & 38.5 & MAGIC & \cite{Albert2006b}\\
1ES 1101-232 & 0.186 & 2.94 & 5.63 & 0.658 & 1.7 & 2.17 & HESS & \cite{Aharonian2007e}\\
1ES 0347-121 & 0.188 & 3.1 & 4.52 & 0.632 & 1.8 & 1.94 & HESS & \cite{Aharonian2007f}\\
1ES 1011+496 & 0.212 & 4.0 & 3.20 & 1.33 & 2.5 & 3.52 & MAGIC & \cite{Albert2007d} \\\hline\hline  
\end{tabular}                          
\end{table}
\begin{figure}[!!h!tb]
\begin{minipage}[c]{1.\linewidth}
\centering
\includegraphics*[width=60mm, viewport = 45 45 322 302]{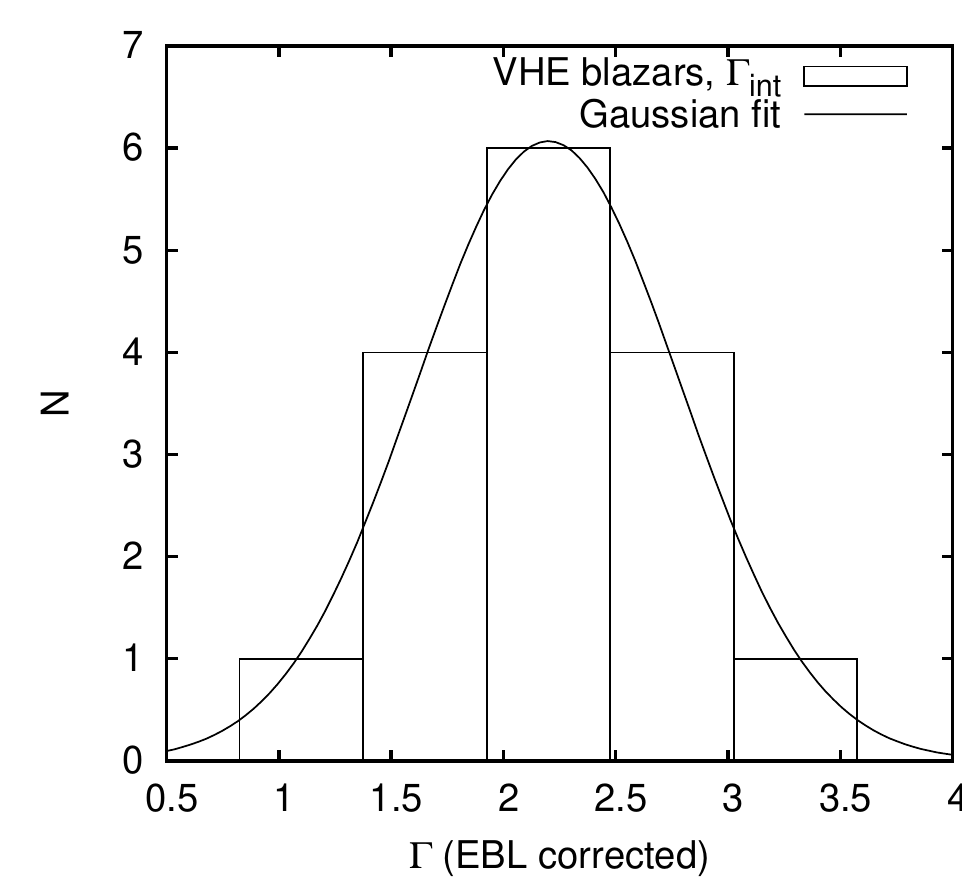}
\end{minipage}
\begin{minipage}[c]{1.0\linewidth}
\centering
\includegraphics*[width=61mm, viewport = 45 45 320 300]{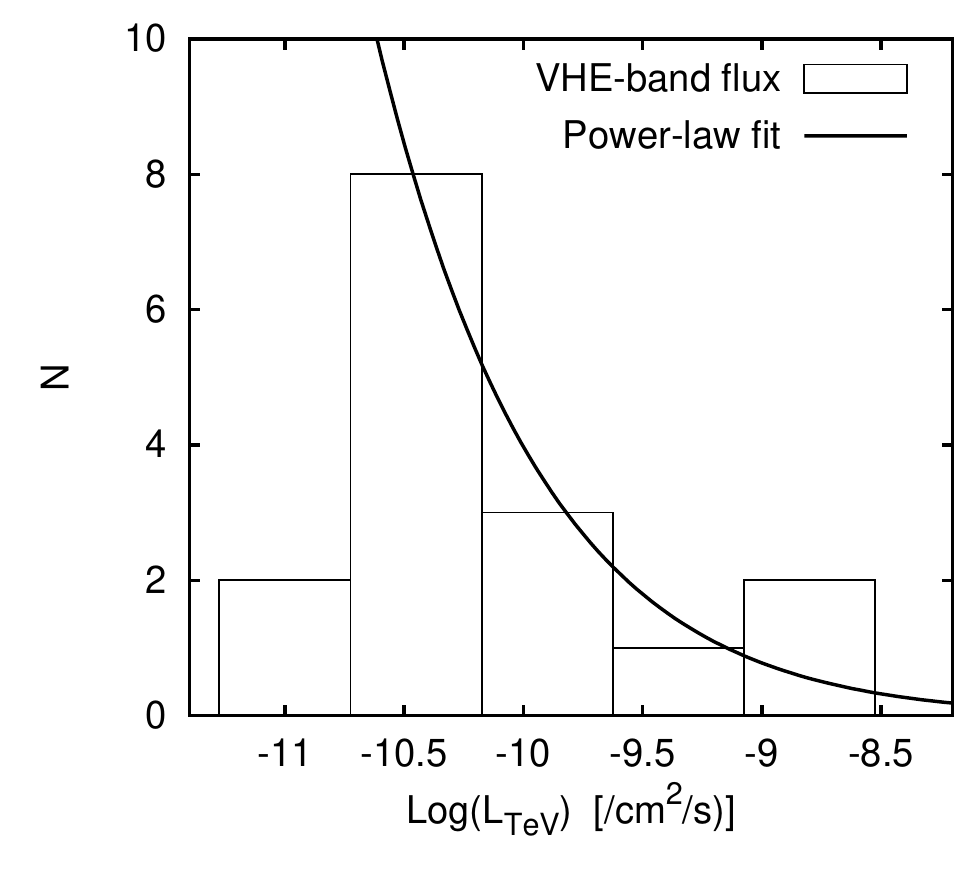}
\caption{Distribution of the various properties of the VHE blazar sample. \textbf{Left:} The distribution of the intrinsic $\Gamma$ (EBL absorption corrected) and the Gaussian fit (with mean at 2.2) to it. \textbf{Right:} The EBL corrected flux ($0.2$\,TeV\,$<$\,E\,$<10$\,TeV), and the powerlaw fit (with slope of 15.481) used for simulations.}
\end{minipage}
\label{ObsDistribution}
\end{figure}

This might be the result of the actual distribution of the intrinsic properties of the parent blazar sample, which this small sample of detected sources correctly reflects. However, it might also be due to selection effects, since all these sources were carefully selected for pointed observation with ACT instruments based on certain criteria, such as described in \cite{Costamante2002}, and not the result of an unbiased all sky survey. In this work, a preliminary attempt has been made to explore the first of the two possibilities mentioned above.

\section{Simulating VHE blazars}
Neglecting at first the selection effects involved in choosing blazars for VHE observations with ACT, various parent samples are simulated. The properties used to characterize each parent sample are the distribution of the intrinsic photon index, the deabsorbed band flux (as a proxy for the intrinsic luminosity), and the redshift distribution (i.e. the density evolution). The deabsorbed photon index ($\Gamma_{int}$) distribution is well described by a Gaussian, Figure \ref{ObsDistribution} left panel, and the band flux (L$_{TeV}$, between $0.2$\,TeV to $10$\,TeV) is fitted with a power-law in $N$ versus Log(L$_{TeV}$). \begin{figure}[!!h!tb]
\begin{minipage}[c]{1.\linewidth}
\centering
\includegraphics*[height=.33\textheight, viewport = 50 55 295 295]{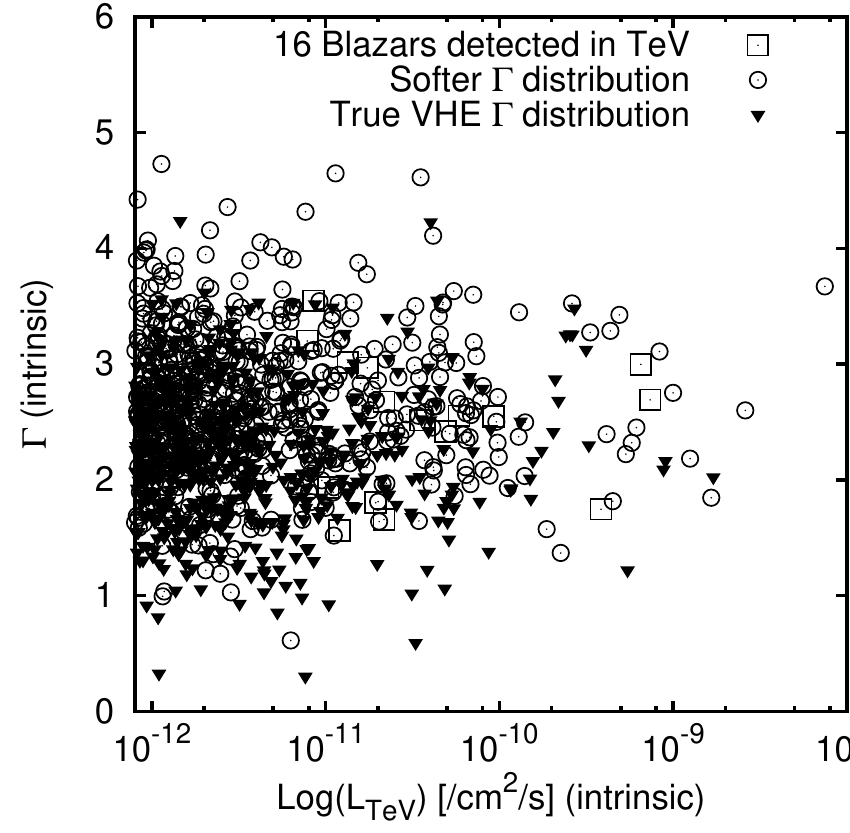}
\end{minipage}
\hspace{0.0cm}
\begin{minipage}[c]{1.0\linewidth}
\centering
\includegraphics*[height=.33\textheight, viewport =  50 50 295 295]{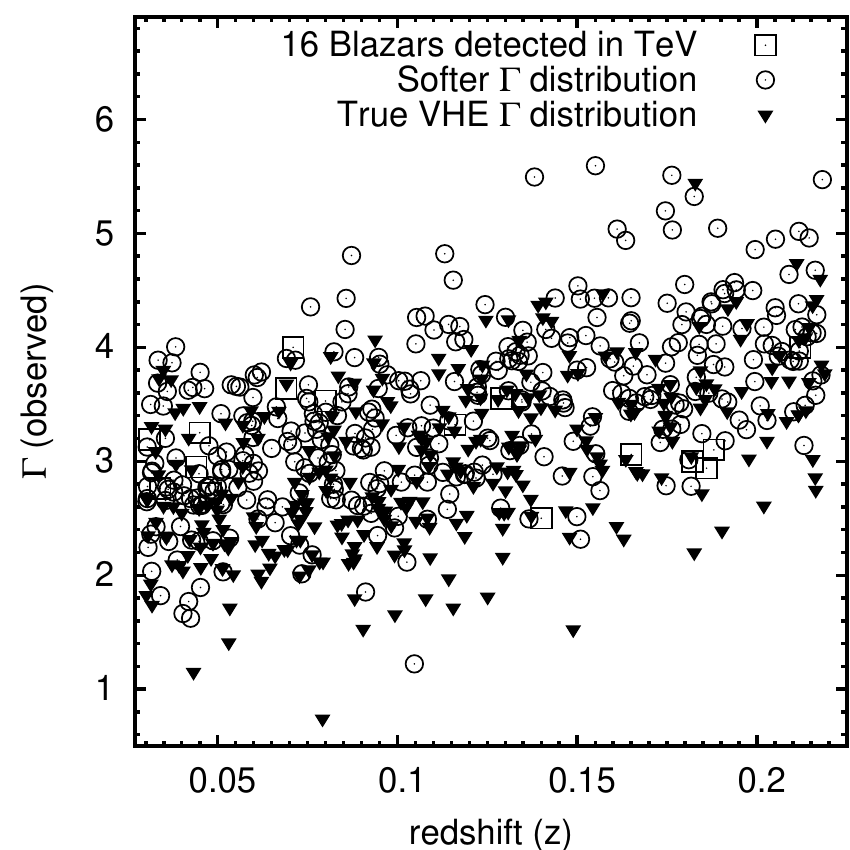}
\caption{Results of simulation of a 1000 blazars; squares (in both panels) are the true VHE blazars. \textbf{Left:} The intrinsic $\Gamma$ versus the deabsorbed band luminosity ($0.2$\,TeV\,$<$\,E\,$<10$\,TeV) of the simulated parent population (limited to deabsorbed L$_{TeV}>10^{-12} /$\,cm\,$^2/$\,s). Inverted triangles denote the parent sample with the luminosity and $\Gamma$ distribution derived from the 16 VHE sources (Figure \ref{ObsDistribution}). Circles denote the parent population with the mean intrinsic-$\Gamma$ shifted to $2.7$. \textbf{Right:} The simulated sources that survived the HESS sensitivity cut (368 circles and 309 triangles) on the EBL-absorbed Flux (above $0.2$\,TeV). There is a clear softening trend of the observed $\Gamma$ with redshift, in both cases.}
\end{minipage}
\label{Simulation1n2}
\end{figure}
Each simulated blazar is uniquely represented by a combination of z, $\Gamma_{int}$ and L$_{TeV}$, where the value of each quantity is obtained from simulations of particular functional forms of the various distributions as described below. For all sets of simulations an uniform z-distribution is taken. Blazars in the simulated samples are individually corrected for intergalactic attenuation due to the EBL and a detector sensitivity cut (for simplicity only a cut to HESS sensitivity is considered here) is applied to get the subsample of blazars that could be detected by the current generation of ACTs. The resulting subsamples are then compared to the true VHE blazar sample.

Three sets of simulations were considered:

1) For the first set, the functional forms shown in Figure \ref{ObsDistribution} were used for the L$_{TeV}$ and $\Gamma_{int}$ distributions to simulate 1000 artificial blazars (inverted triangles in Figure~\ref{Simulation1n2}), which were considered as the parent sample. The absorbed spectra was calculated, using the relevant optical depths calculated from the EBL model taken from \cite{Aharonian2006a}. The integral flux in the energy range $0.2$\,TeV to $10$\,TeV of the resulting spectra was found. Sources with the integral flux greater $10^{-12}$/cm$^2$/s were considered to be detectable with the present generation of Cherenkov telescopes. The fact that no value in column $5$ of Table~\ref{rankingtable} is lower than this value, demonstrates that this is indeed an accurate description of the detection threshold. Results are shown in the right panel of Figure~\ref{Simulation1n2}, as inverted triangles.
\begin{figure}[!!h!!bt]
\begin{minipage}[c]{1.\linewidth}
\centering
\includegraphics*[height=.33\textheight, viewport = 50 55 295 295]{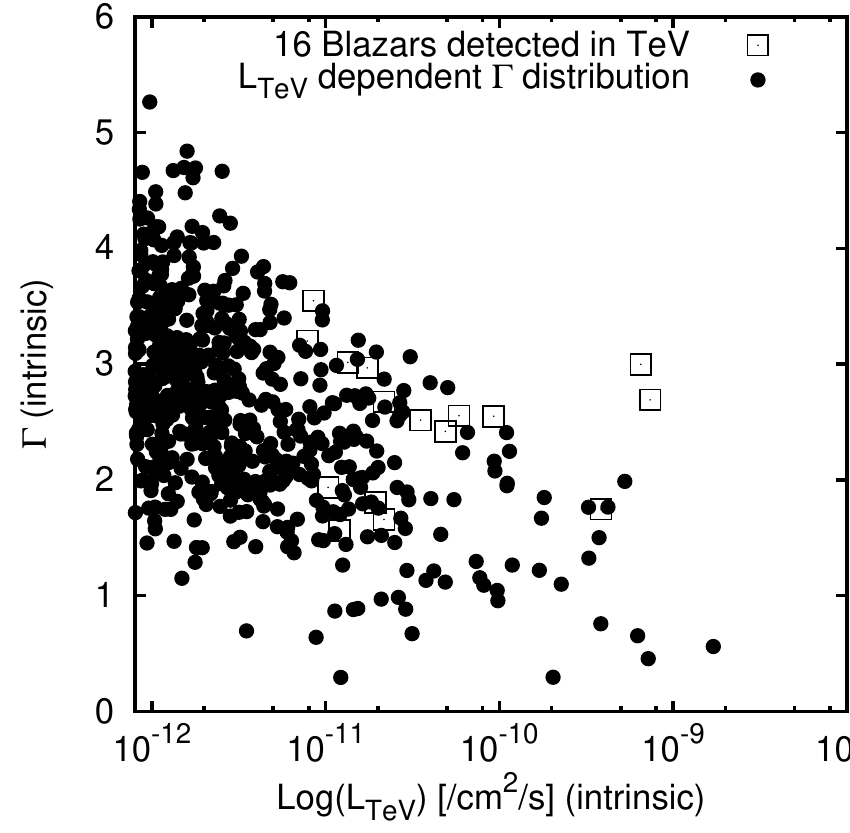}
\end{minipage}
\hspace{0.0cm}
\begin{minipage}[c]{1.0\linewidth}
\centering
\includegraphics*[height=.33\textheight, viewport =  50 50 295 295]{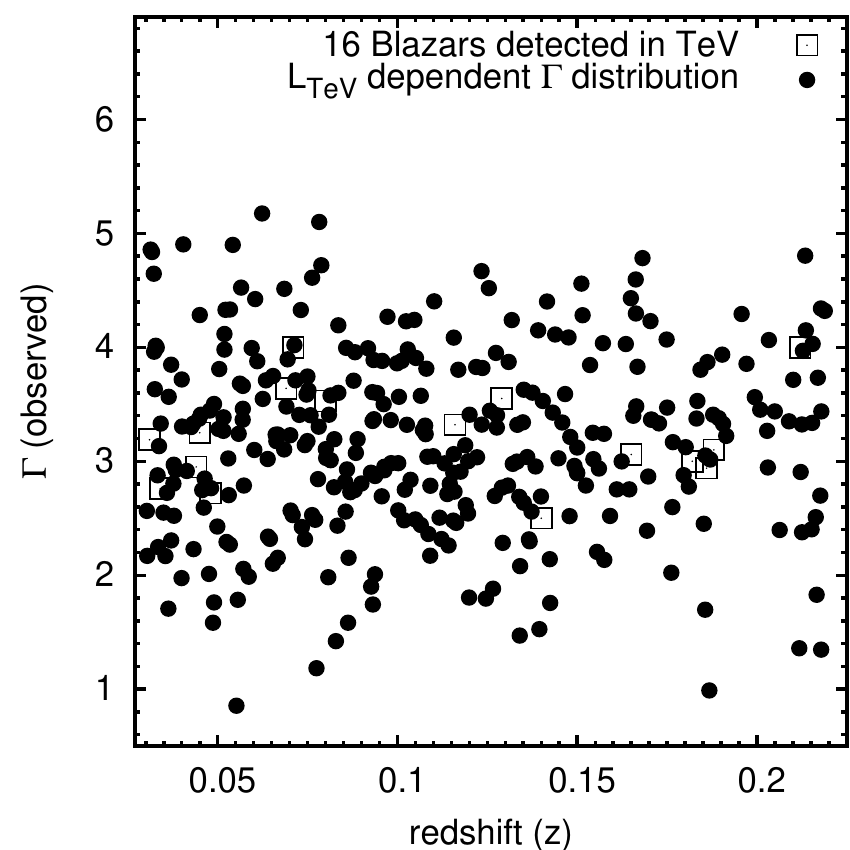}
\caption{The simulation of 1000 blazars (filled circles); squares are the true VHE blazars. \textbf{Left:} The intrinsic $\Gamma$ versus band luminosity of the parent sample, clearly seen is the inverse relation between Log(L$_{TeV}$) and $\Gamma$ assumed for this run. \textbf{Right:} The simulated sources that survived the HESS sensitivity cut (335 in total) on the EBL absorbed Flux (above $0.2$\,TeV). There appears to be no hint of a spectral softening with z for this simulation.}
\end{minipage}
\label{Simulation3}
\end{figure}

2) The parent blazar sample for the second set was chosen with the same L$_{TeV}$ distribution as above but with the mean of $\Gamma_{int}$ distribution softened to $2.7$ instead. The other parameters for the Gaussian fit to $\Gamma_{int}$ were not modified. The same recipe was followed for extracting the EBL-absorbed spectra and the get the sub sample of sources that can be detected (see Figure \ref{Simulation1n2}, circles).

3) The parent sample for the last set of simulations was chosen with the same L$_{TeV}$ as above but with the mean of $\Gamma_{int}$ distribution chosen according to a simple inverse linear relation between Log(L$_{TeV}$) and $\Gamma_{int}$, representing the so called ``blazar sequence''. According to the blazar sequence, \cite{Fossati1998}, \cite{Ghisellini1998} and references therein, blazars with higher luminosity (over all $\lambda$, i.e. the bolometric luminosity) have the synchrotron peak at lower energies, and have a softer spectral index. Higher overall luminosity with softer spectral index roughly translates to lower band flux in $0.2$\,TeV to $10$\,TeV, since brighter sources with softer spectra drop sharply at high energies (i.e. when extended to VHE) whereas fainter sources with harder spectra extend further into VHE before falling off. Hence the relation seen in Figure \ref{Simulation3}, left panel. To get the subsample of blazars that can be detected with ACTs the same procedure as the above cases was applied.

\section{Results and Discussion}
The first two simulations show a softening trend of the observed photon index with redshift as seen in Figure \ref{Simulation1n2} (for both the true distribution as well as a softer range of $\Gamma_{int}$) contrary to the lack of any similar trend in the redshift-$\Gamma$ relation of the 16 known VHE blazars. The third simulation (analogous to the so called ``blazar sequence'') does not show any such trends.

It should however be pointed out that the blazar sequence is derived from a highly (observationally) biased sample, and might not reflect a true systematic trend in the physical properties of AGN. Also, note that the instrumental bias in the sample of the VHE blazars is not uniform since the measurements were taken with different instruments having different sensitivities. An additional caveat regarding the simulations is that the proxy used for luminosity, i.e. the band flux has a z dependence and hence is not a completely independent intrinsic parameter for the simulated blazars. Simulating the intrinsic band luminosity instead of the band flux produces identical results, and will be the parameter used in future simulations to be published elsewhere. Despite these shortcomings the results of these preliminary simulations are encouraging, as it provides evidence that the presence or the lack of a spectral softening trend with z, could be the result of a peculiar parent population, though the instrumental bias has to be carefully studied.

\section{Conclusion}
We tentatively conclude that the observational bias and instrumental-sensitivity limitations of Cherenkov telescopes, are responsible for cancelling out the expected trend of spectral softening (due to EBL attenuation) with z, in blazars detected in VHE $\gamma$-rays. More detailed simulations and statistical tests will be done to make quantitative conclusions about the current VHE observations, and to make predictions about the trends that might appear with a larger VHE blazar sample.
 

\begin{theacknowledgments}
This work is supported by the DFG through SFB 439.
\end{theacknowledgments}

\end{document}